\documentclass[aps,prl,twocolumn,floatfix]{revtex4}
\usepackage{amssymb}
\usepackage{amsmath}
\usepackage{graphicx}

\begin{document}

\title{Crossing points of nodal lines in topological semimetals  and Fermi surface of ZrSiS}

\author{G.~P.~Mikitik}
\affiliation{B.~Verkin Institute for Low Temperature Physics \&
Engineering, Ukrainian Academy of Sciences, Kharkiv 61103,
Ukraine}

\author{Yu.~V.~Sharlai}
\affiliation{B.~Verkin Institute for Low Temperature Physics \&
Engineering, Ukrainian Academy of Sciences, Kharkiv 61103,
Ukraine}

\begin{abstract}
We investigate the electron spectra, Fermi surfaces and their characteristics near crossing points of two band-contact lines in nodal-line semimetals. In particular, the extremal cross-sectional areas, and the appropriate cyclotron masses are calculated. We also find the phase of the quantum oscillations associated with the electron orbits near the crossing point. The analysis of all these quantities is carried out both without and with consideration of the spin-orbit interaction. To illustrate the obtained results, we apply them to ZrSiS in which the crossing of the nodal lines occurs.
\end{abstract}

\maketitle

\section{Introduction}

Band-contact lines (nodal lines) along which two electron energy bands touch in a Brillouin zone, are widespread in crystals \cite{herring,m-sh14,kim,fang}. For example,
such contacts of the bands occur in Bernal graphite \cite{graphite}, beryllium \cite{beryl,beryl1}, magnesium \cite{beryl1}, aluminium \cite{al}, LaRhIn$_5$ \cite{prl04}, and in the bulk Rashba semiconductors BiTeI and BiTeCl (see, e.g., \cite{m-sh19}). Besides, the band-contact lines exist in all the topological nodal-line semimetals which have attracted a lot of attention in recent years  \cite{m-sh19,armit,bernevig,gao,weng-r,fang-r}. It is necessary to emphasize that the band-degeneracy energy $\varepsilon_d$ at which the bands touch generally is not constant, and $\varepsilon_d$ changes along a band-contact line in the interval between its minimum  $\varepsilon_{min}$ and maximum $\varepsilon_{max}$ values. The distinctive feature of the nodal-line semimetals is that the
difference $\varepsilon_{max}-\varepsilon_{min}$ is small as compared to the characteristic scale $\varepsilon_0 \sim 1-10$ eV of the electron band structure in crystals, and the chemical potential $\zeta$ of the charge carriers does not lie far away from the mean energy $\varepsilon_d^0 \equiv (\varepsilon_{max}+ \varepsilon_{min})/2$ of the line.

\begin{figure}[tbp] 
 \centering  \vspace{+9 pt}
\includegraphics[scale=.8]{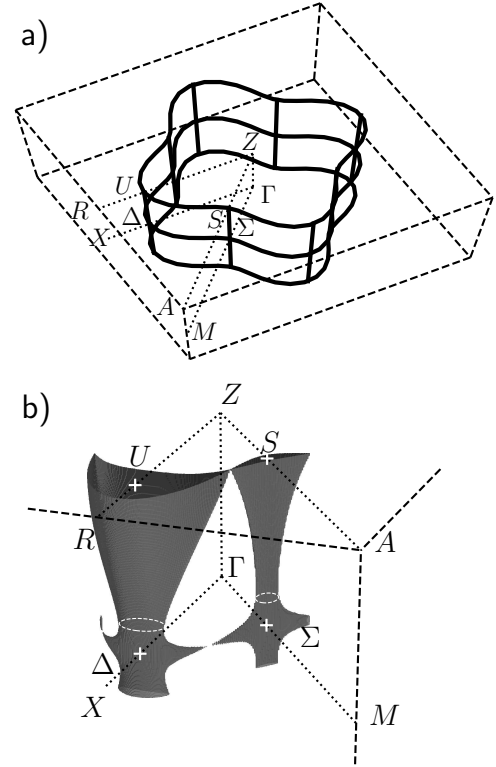}
\caption{\label{fig1} a) The Brillouin zone (the dashed lines), the symmetry directions (the dotted lines), the band-contact lines represented schematically (the solid lines), and the nonequivalent crossing points $\Delta$, $\Sigma$, $U$, $S$ in ZrSiS. b) The part of the Fermi surface in ZrSiS shown schematically at  $\varepsilon_{cr}^{U} > \varepsilon_{cr}^{\Delta}>\zeta > \varepsilon_{cr}^{\Sigma} > \varepsilon_{cr}^{S}$. The crosses mark the points $\Delta$, $\Sigma$, $U$, $S$, and the extremal orbits on the necks of this surface are sketched by the white dashed circles. The realistic Fermi-surface is depicted in Fig.~1 of Ref.~\cite{pez}.
 } \end{figure}   

In a plane  perpendicular to an {\it isolated} band-contact line, the gap between the two contacting bands is proportional to the deviation of the quasi-momentum ${\bf p}$ from the line, i.e., near such lines the spectrum has the Dirac form \cite{m-sh16,m-sh18}. However, the band-contact lines can cross each other at some points in symmetry axes of crystals, and in the vicinity of the crossing point, the electron spectrum essentially  changes. Below we consider the simplest situation when only two band-contact lines cross. Such crossings can occur in twofold or fourfold symmetry axes. In partucular, the crossing of this type takes place in Mackay-Terrones crystals \cite{weng}, ZrB$_2$ \cite{lou18,wang18}, V$_3$Si \cite{gork,step}, and in the ZrSiS-family of the nodal-line semimetals \cite{schoop,pez,fu19,chen17,hosen,pan,delft,guo19}. In this paper, to illustrate the obtained general results, we shall apply them to ZrSiS in which the energies $\varepsilon_{cr}^i$ of the crossing points lie near the the chemical potential $\zeta$ (the index $i$ marks these points).

In the Brillouin zone of ZrSiS, the nodal lines form a ``cage'' with four nonequivalent crossing points $\Delta$, $\Sigma$, $U$, $S$ lying  in the axes $\Gamma$-X, $\Gamma$-M, Z-R, Z-A, respectively; see Fig.~\ref{fig1}. The Fermi surface of ZrSiS can be qualitatively described as a connected net of electron and hole tube-like surfaces, and of self-intersecting surfaces composed of electron and hole parts. Each of these surfaces encloses a portion of the nodal line between two crossing points, see Fig.~\ref{fig1}. When the chemical potential changes and passes one of the energies $\varepsilon_{cr}^{i}$, a part of the electron (hole) ``tubes'' evolves into the  self-intersecting surfaces or vice versa.
Since the energies  $\varepsilon_{cr}^{\Delta}$ and $\varepsilon_{cr}^{\Sigma}$ are close to each other, the Fermi surface is very sensitive to the Fermi-level position relative to these energies. In particular, a small doping can noticeably change a  part of the Fermi surface. This sensitivity also leads to somewhat different Fermi surfaces obtained in the band structure calculations for ZrSiS \cite{pez,fu19}. As a result, it is difficult to identify the extremal cross sections associated with the small frequency oscillations observed in the resistivity, magnetization, and the thermoelectric power of the ZrSiS family of the nodal-line semimetals \cite{delft,guo19,Ali1,wang1,singha,kumar,Hu,Hu2,Hu1,Hu18,matus,muller}.

In Ref.~\cite{step}, a ${\bf k}\cdot {\bf p}$ model was suggested that describes the electron energy spectrum in the vicinity of a crossing point of two band-contact lines. Using this general model, in Sec. II we classify the Fermi surfaces and calculate the  extremal cross-sectional areas and the cyclotron masses near the crossing points in the nodal-line semimetals. In Sec.~III we consider the effect of the weak spin-orbit interaction on the quantities considered in Sec.~II.  We also calculate the phase $\phi$ of the quantum oscillations for the electron orbits near the crossing points. Without the spin-orbit coupling, this $\phi$ is determined by the Berry phase  and has the unique value \cite{prl}. However, the spin-orbit interaction {\it together with} proximity of the orbit to the crossing point can noticeably change the phase of the oscillations. All these results can be useful in analyzing experimental data of various oscillation experiments. Relying on the findings of Secs. II and III,  we discuss the case of ZrSiS  in Sec.~IV. Conclusions are presented in Sec.~V.

\section{Fermi surface and its characteristics near crossing points}

\subsection{Spectrum of electrons and types of Fermi surface}

\begin{figure}[tbp] 
 \centering  \vspace{+9 pt}
\includegraphics[scale=1.0]{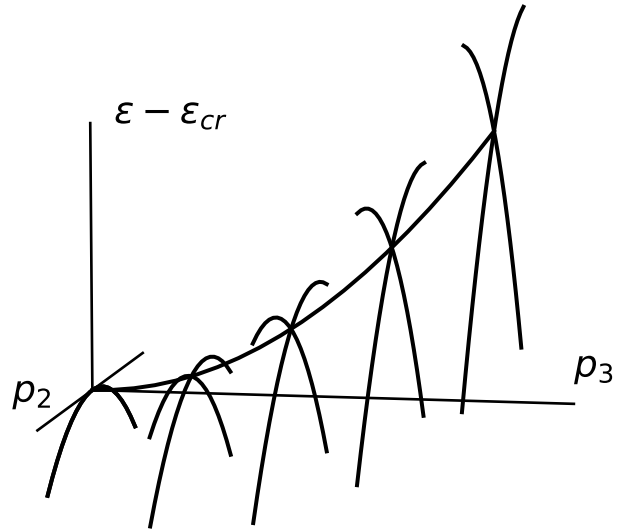}
\caption{\label{fig2} The dispersion laws $\varepsilon_{c,v}(p_2)$ of the two electron energy bands, described by Eqs.~(\ref{1}) and (\ref{2}), at $p_1=0$ and various fixed values of $p_3$. The band-contact line ($p_3=0$) gradually evolves into the Dirac spectrum when $p_3$ increases, i.e., when the plane perpendicular to the other band-contact line ($p_2=0$) moves away from the crossing point ($p_2=p_3=0$) of these lines. For definiteness,  $B_2<0$, $B_3>0$, and $B_2'=B_3'=0$ here.
 } \end{figure}   

Neglecting the spin-orbit interaction, the electron energy spectrum for the two bands ``$c$'' and ``$v$'' in the vicinity of the crossing point with the energy $\varepsilon_{cr}$ has the form \cite{step}:
 \begin{eqnarray}\label{1}
 \varepsilon_{c,v}({\bf p})\!\!&=&\!\varepsilon_{cr}+a p_1+B_2 p_2^2 + B_3 p_3^2+E_{c,v}({\bf p}), \\
 E_{c,v}({\bf p})\!\!&=&\!\pm \left[(a'p_1+B_2'p_2^2+B_3'p_3^2)^2+ \beta^2p_2^2p_3^2\right]^{1/2}\!\!\!, \label{2}
 \end{eqnarray}
where the $p_1$ axis coincides with the symmetry axis in which the crossing point is located; the axes $p_2$ and $p_3$ are along the tangents to the band-contact lines at their crossing point;
all the quasi-momenta $p_1$, $p_2$, $p_3$ are measured from this point; $a$, $a'$, $B_i$, $B_i'$, $\beta$  are constant parameters of the spectrum \cite{com1}; see Fig.~\ref{fig2}. In particular, in the case of ZrSiS  the $p_3$ axis is parallel to $\Gamma$-Z direction, whereas the coordinate $p_1$ is measured along the symmetry axes $\Gamma$-X, $\Gamma$-M, Z-R, Z-A for the points $\Delta$, $\Sigma$, $U$, $S$, respectively. The spectrum described by Eqs.~(\ref{1}) and (\ref{2}) is valid  when the energy of the charge carriers is close to $\varepsilon_{cr}$, $|\zeta- \varepsilon_{cr}| \ll \varepsilon_0$. In the nodal-line semimetals this restriction becomes more rigid, $|\zeta- \varepsilon_{cr}| \ll \varepsilon_{max} - \varepsilon_{min}$.

\begin{figure}[tbp] 
 \centering  \vspace{+9 pt}
\includegraphics[scale=1.0]{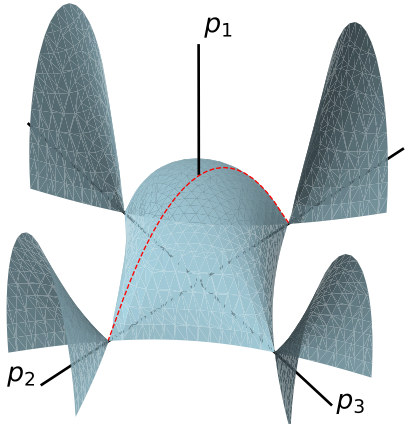}
\caption{\label{fig3} The Fermi surface near a crossing point for $1>\lambda>0$ and $(\zeta-\varepsilon_{cr})B_3>0$ where $\lambda \equiv 4B_2B_3/\beta^2$, Eq.~(\ref{7}). For clarity, only a half of the Fermi surface (at $p_1>0$) is shown. There are no necks on the Fermi surface. The orbit corresponding to $S_{3,max}$ is shown by the dashed line.
 } \end{figure}

\begin{figure}[tbp] 
 \centering  \vspace{+9 pt}
\includegraphics[scale=1.0]{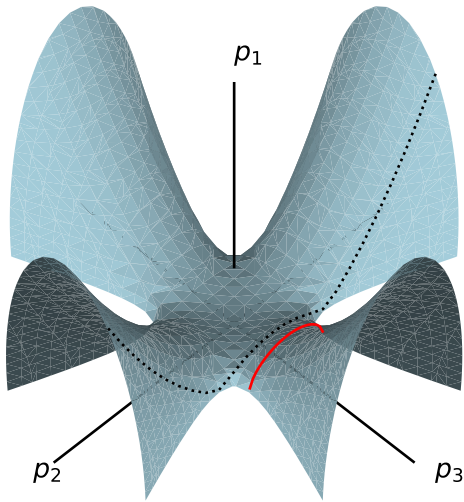}
\caption{\label{fig4} The Fermi surface (at $p_1>0$) near a crossing point for $1>\lambda>0$, but at $B_3(\zeta-\varepsilon_{cr})<0$.
There are four necks on the Fermi surface. The extremal orbit on
one of the necks is shown by the solid line while the self-intersecting orbit is marked by the dotted line.
 } \end{figure}

\begin{figure}[tbp] 
 \centering  \vspace{+9 pt}
\includegraphics[scale=1.0]{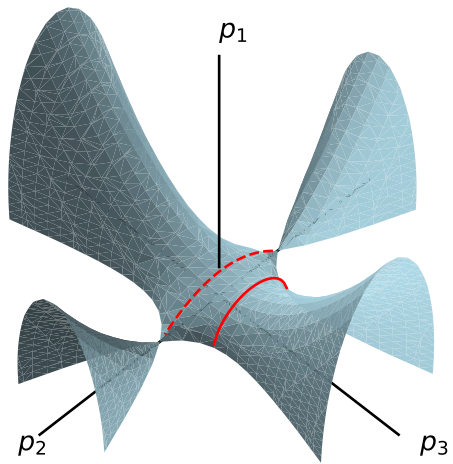}
\caption{\label{fig5} The Fermi surface (at $p_1>0$) near a crossing point in the case of $\lambda<0$. There are two necks on this  surface near one of the two band-contact lines which coincide with the $p_2$ and $p_3$ axes (for definiteness, the case $B_3(\zeta-\varepsilon_{cr})<0$ is shown here).  If the sign of $(\zeta-\varepsilon_{cr})$ changes, the necks occur near the other band-contact line. The extremal orbit in one of the necks is shown by the solid line, while the orbit corresponding to the maximum cross section is marked by the dashed line.
 } \end{figure}

The nodal lines are determined by the condition $E_{c,v}({\bf p})=0$ which yields the two crossing lines: $p_2=0$, $p_1=-B_3'p_3^2/a'$ and $p_3=0$, $p_1=-B_2'p_2^2/a'$. Although the parameters $B_2'$ and $B_3'$ may be sufficiently large, nonzero values of $B_2'$ and $B_3'$ have no effect on the cross-sectional areas and cyclotron masses
given below and on the topology of the Fermi surface near a crossing point. For this reason, we set $B_2'=B_3'=0$ in our subsequent analysis. To imagine the situation with nonzero values of $B_2'$ and $B_3'$, one should ``bend''  the plane $p_2$-$p_3$ in Figs.~\ref{fig3}-\ref{fig5}.   As to the parameter $a$, it  determines the tilt of the spectrum along the appropriate symmetry axis. Since there is no visible tilt for ZrSiS family of the semimetals  \cite{schoop,pez,fu19,chen17,hosen,pan,delft,guo19}, we assume below that $a\equiv 0$. This assumption simplifies the subsequent formulas while not imposing fundamental restrictions on the results. It is also worth noting that according to Eq.~(\ref{2}), the band degeneracy at the crossing point is lifted linearly in $p_1$ and quadratically for all directions in the $p_2-p_3$ plane except  the band-contact lines. This unusual dispersion law is due to that the familiar Dirac spectrum in the planes perpendicular to a band-contact line is not compatible with the crossing. However, formula (\ref{2}) shows that as the plane perpendicular to one of the crossing lines moves away from the crossing point, the Dirac spectrum is gradually restored, Fig.~\ref{fig2}.

Possible types of the Fermi surface near the crossing points in the nodal-line semimetals are specified by the signs of the product $B_2B_3$ and of $\zeta-\varepsilon_{cr}$. These types are presented in Figs.~\ref{fig3}-\ref{fig5}. They differ in the number of ``necks'' of the Fermi surface in the vicinity of the crossing point. In Figs.~\ref{fig3} and \ref{fig4} we show the case $B_2B_3>0$. If $B_3(\zeta-\varepsilon_{cr}) >0$, see Fig.~\ref{fig3}, the necks are absent, and four self-intersecting surfaces ``emerge from'' the central region containing the crossing point. At the points of the self intersection, the hole and electron pockets touch. However, if one takes into account the weak spin-orbit interaction (see Sec.~III), a small gap appears between the pockets. This means that at the magnetic field $H$ directed along the $p_3$ (or $p_2$) axis, the only extremal cross section passes through the crossing point, and it is the maximal cross section of the central pocket. At $B_3(\zeta-\varepsilon_{cr}) <0$, see Fig.~\ref{fig4}, there is a neck in each ``tube'' emerging from the cental region. In other words, among the cross sections produced by the planes $p_3=$const. (or $p_2=$const.), the minimal one exists in each tube, and this cross section does not pass through the crossing point. When $B_2B_3<0$,  only one type of the Fermi surfaces  is possible. In this case the number of the necks is equal to two, Fig.~\ref{fig5}. If $B_3(\zeta-\varepsilon_{cr})<0$, these two necks occur in the Fermi-surface tubes enclosing the $p_3$ axis, whereas at $B_3(\zeta-\varepsilon_{cr})>0$, the necks belong to the tubes enclosing the $p_2$ axis. In other words, to imagine the case $B_3(\zeta-\varepsilon_{cr})>0$, the Fermi surface shown in Fig.~\ref{fig5} has to be rotated by $\pi/2$, with electrons being replaced by holes and holes by electrons. Besides, Fig.~\ref{fig5} shows that there is also a maximal cross section in the plane $p_3=0$. Finally, it should be noted that at $\lambda\equiv 4B_2B_3/\beta^2>1$ additional two types of the Fermi surface are possible. However, we do not analyze these types here since in the nodal-line semimetals the parameter $\lambda$ is generally small. This is due to the condition $\varepsilon_{max}- \varepsilon_{min}\ll \varepsilon_0$ mentioned in the Introduction, see also Sec.~IV.

Appearance (disappearance) of self-intersecting Fermi surfaces in  nodal-line semimetals with changing $\zeta$ is the electron topological $3\frac{1}{2}$-order transition that takes place near the critical energies $\varepsilon_{max}$ and $\varepsilon_{min}$ \cite{m-sh14,m-sh-jltp}. According to Figs.~\ref{fig3}-\ref{fig5}, such transformations of the Fermi surface can also occur near the crossing points of the nodal lines when the chemical potential passes the appropriate energy $\varepsilon_{cr}$ (e.g., the surface in Fig.~\ref{fig4} transforms into that shown in Fig.~\ref{fig3}). However, the special ${\bf p}$-dependence of the energy bands $\varepsilon_{c,v}({\bf p})$ near the crossing point, Eqs.~(\ref{1}) and (\ref{2}), leads to that this transition is of  the $3$rd kind according to the classification of Lifshitz \cite{lif}. This transition is characterized by a specific dependence of the magnetic susceptibility on the chemical potential $\zeta$ \cite{step}. Below, analyzing the Fermi surfaces near the crossing points, we study their  characteristics  that are measured in oscillation experiments.

\subsection{Cross-sectional areas}

Using Eqs.~(\ref{1}) and (\ref{2}), one can calculate areas $S_3$ of the cross sections of the Fermi surface by the planes perpendicular to the axis $p_3$:
\begin{eqnarray}\label{3}
 S_3(\zeta,p_3)&=&\frac{4|\zeta-\varepsilon_{cr}|^{3/2}}{3a'|B_2|^{1/2}} F_3(\tilde p_3),\\
 F_3(\tilde p_3)&=&x_b\left[ (x_a^2+x_b^2)E(t) -(x_b^2-x_a^2)K(t)\right], \label{4}
 \end{eqnarray}
where $t=x_a/x_b$ is the modulus of the complete elliptic integrals $E(t)$ and $K(t)$ \cite{BE}, $\tilde p_3 \equiv |B_3|^{1/2}p_3/|\zeta-\varepsilon_{cr}|^{1/2}$ is the dimensionless quasi-momentum,
\begin{eqnarray}\label{5}
 x_a&=&\left|\sqrt{\xi_2\xi_{\zeta}-\xi_{\lambda}\tilde p_3^2
+\frac{\tilde p_3^2}{|\lambda|}}-\frac{|\tilde p_3|}{\sqrt{|\lambda|}}\right|, \\
x_b&=&\sqrt{\xi_2\xi_{\zeta}-\xi_{\lambda}\tilde p_3^2
+\frac{\tilde p_3^2}{|\lambda|}}+\frac{|\tilde p_3|}{\sqrt{|\lambda|}}, \label{6}\\
\lambda&\equiv& \frac{4B_2B_3}{\beta^2}, \label{7}
   \end{eqnarray}
$\xi_{\zeta}\equiv {\rm sign}(\zeta-\varepsilon_{cr})$ is equal to $+1$ at $(\zeta-\varepsilon_{cr})>0$ and $-1$ at $(\zeta-\varepsilon_{cr})<0$; similarly, $\xi_2$, $\xi_3$, $\xi_{\lambda}$ are signs of $B_2$, $B_3$, $\lambda$, respectively. The function $F_3(\tilde p_3)$ describes the dimensionless cross-sectional areas.

\begin{figure}[tbp] 
 \centering  \vspace{+9 pt}
\includegraphics[scale=1.0]{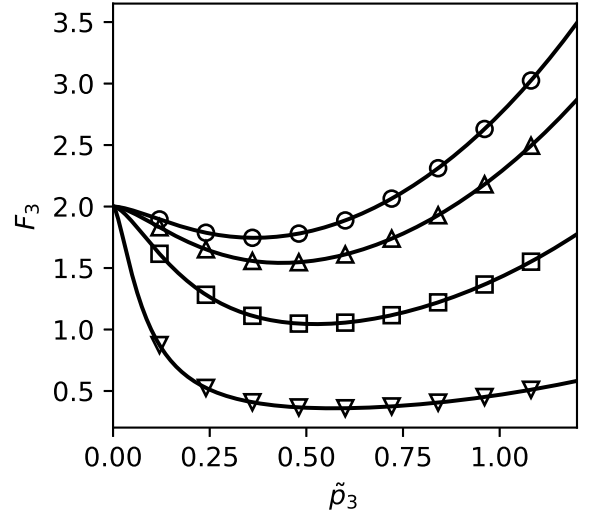}
\caption{\label{fig6} The function $F_3(\tilde p_3)$, Eq.~(\ref{4}), at $B_3(\zeta-\varepsilon_{cr})<0$ for different {\it negative} values of the parameter $\lambda$: $-0.01$ ($\nabla$), $-0.1$ ($\Box$), $-0.3$ ($\triangle$), $-0.5$ ($\bigcirc$). Here $\tilde p_3 \equiv  |B_3|^{1/2}p_3/|\zeta-\varepsilon_{cr}|^{1/2}$.
 } \end{figure}   

\begin{figure}[tbp] 
 \centering  \vspace{+9 pt}
\includegraphics[scale=1.0]{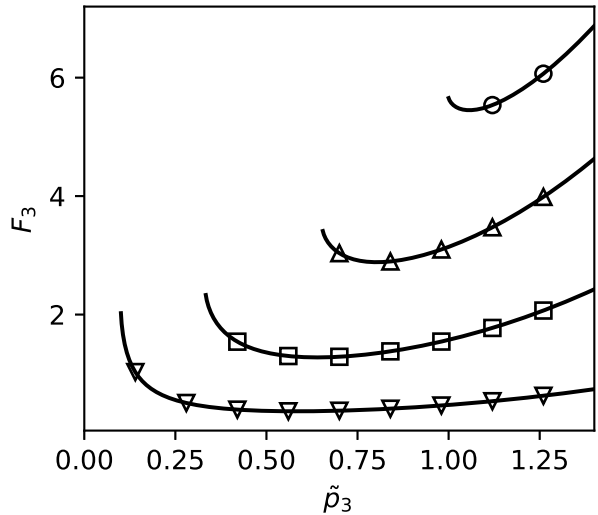}
\caption{\label{fig7} The function $F_3(\tilde p_3)$, Eq.~(\ref{4}), at $B_3(\zeta-\varepsilon_{cr})<0$ for different {\it positive} values of the parameter $\lambda$: $0.01$ ($\nabla$), $0.1$ ($\Box$), $0.3$ ($\triangle$), $0.5$ ($\bigcirc$). Here $\tilde p_3 \equiv  |B_3|^{1/2}p_3/|\zeta-\varepsilon_{cr}|^{1/2}$. The solid lines terminate when the closed orbit on the neck transforms into the self-intersecting orbit shown in Fig.~\ref{fig4}.
 } \end{figure}

The dependences of the cross-sectional area $S_3$ on $p_3$ are shown in Figs.~\ref{fig6} and \ref{fig7} for negative and positive values of $\lambda$ and under the condition $B_3(\zeta-\varepsilon_{cr})<0$ ensuring existence of the necks on the Fermi-surface part enclosing the $p_3$ axis. The minimum $S_{3,min}$ of the function $S_3(p_3)$ just corresponds to the extremal orbit on such a neck.
As was mentioned above, the nodal-line semimetals are characterized by small values of $|\lambda|$. In this situation formulas (\ref{4})-(\ref{6}) can be simplified in the vicinity of the neck  (i.e., at $\tilde p_3 \sim 1$) as follows:
\begin{eqnarray}\label{8}
 x_a\!\!&\approx&\!\frac{|\lambda|^{1/2}}{2}\frac{1+\tilde p_3^2}{|\tilde p_3|},\ \ \
x_b\approx\frac{2|\tilde p_3|}{|\lambda|^{1/2}},\ \ \ t=\frac{x_a}{x_b}\ll 1,~~\nonumber \\
E(t)\!\!&\approx&\!\frac{\pi}{2}\left(1-\frac{t^2}{4}\right), \ \ \ K(t)\approx \frac{\pi}{2}\left(1+\frac{t^2}{4}\right),  \nonumber \\
 F_3(\tilde p_3)\!\!&\approx&\!\frac{3\pi}{4}x_b x_a^2\approx \frac{3\pi|\lambda|^{1/2}}{8}\frac{(1+\tilde p_3^2)^2}{|\tilde p_3|}.
   \end{eqnarray}
The minimum of the function $F_3(\tilde p_3)$ described by  Eq.~(\ref{8}) is reached at $\tilde p_{3,min}=1/\sqrt{3}$ and is equal to
\begin{eqnarray}\label{9}
 F_3(\tilde p_{3,min})=\frac{2}{\sqrt{3}}\pi|\lambda|^{1/2}.
 \end{eqnarray}

For $\lambda<0$ (Figs.~\ref{fig5} and \ref{fig6}), apart from $S_{3,min}$, there is also a maximal cross section at $\tilde p_3=0$, and formulas (\ref{4})-(\ref{6}) give $F_3(0)=2$ for this $S_{3,max}$. It also follows from these formulas that the ratio $S_{3,min}/S_{3,max}$ is a  function of $\lambda$ only, and therefore the parameter $\lambda$ can be found if the oscillation frequencies corresponding to both these cross sections are detected.
Our analysis shows that the formula
 \begin{eqnarray}\label{10}
 \frac{S_{3,min}}{S_{3,max}}=\frac{F_3(\tilde p_{3,min})}{2} \approx \frac{1-\exp(-\pi|\lambda|^{1/2}/\sqrt{3})}{1+2.32|\lambda|},
  \end{eqnarray}
sufficiently well describes the ratio at all negative $\lambda$ with $|\lambda|\lesssim 3$.

For the case $\lambda>0$, finite cross-sectional areas $S_3$ exist only at $\tilde p_3^2 \ge \lambda/(1-\lambda)$, Figs.~\ref{fig4} and \ref{fig7}. At $\tilde p_{3,s-i}=\sqrt{\lambda}/(1-\lambda)^{1/2}$ the self-intersecting trajectory occurs, and $F_3(\tilde p_{3,s-i})=2/(1-\lambda)^{3/2}$. Interestingly, for $1/2 \lesssim \lambda <1$, $\tilde p_{3,min}$ and $F_3(\tilde p_{3,min})$ are close to $\tilde p_{3,s-i}$ and $F_3(\tilde p_{3,s-i})$, respectively.

It is also worth noting that according to  Figs.~\ref{fig6} and \ref{fig7}, the function $S_3(p_3)$ is sufficiently ``flat'' near $p_{3,min}$ at small $\lambda$. Hence, if the magnetic field is directed at not-too-large angle $\theta$ to the $p_3$ axis, the minimal cross-sectional area corresponding to the extremal orbit on the neck has the form  $S_{3,min}(\theta)\approx S_{3,min}(0)/\cos\theta$ characteristic of two-dimensional Fermi surfaces.

The minimum of the function $S_3(p_3)$  disappears  when $B_3(\zeta-\varepsilon_{cr})>0$. At $\lambda>0$, the only extremal cross section is the central one with $F_3(0)=2$, Fig.~\ref{fig3}, whereas at $\lambda<0$ the extremal points of $S_3(p_3)$ are absent at all (this conclusion is clear from Fig.~\ref{fig5} if one considers the cross sections by the planes $p_2=$const.).

\subsection{Cyclotron masses}

Using formula (\ref{3}) and taking into account that $\partial S(\zeta,p_3)/\partial p_3=0$ at $p_3=p_{3,ex}$ where $p_{3,ex}$ corresponds to an extremal cross section (i.e., $p_{3,ex}=p_{3,min}$ or $0$), we find the cyclotron masses $m_*=(1/2\pi)\partial S(\zeta,p_3)/\partial \zeta$ for the extremal  cross sections,
\begin{eqnarray}\label{11}
 |m_{*,ex}|=\frac{3S_3(\zeta,p_{3,ex})}{4\pi |\zeta-\varepsilon_{cr}|}.
  \end{eqnarray}
Formula (\ref{11}) enables one to find $\zeta-\varepsilon_{cr}$ if $S_{3,ex}=S_3(\zeta,p_{3,ex})$ and $m_{*,ex}$ are known from an experiment.

\subsection{Phase of quantum oscillations}

It is known \cite{prl} that a band-contact line can lead to the phase  shift of the quantum oscillation as compared to the familiar case \cite{Sh}. Consider, e.g., the oscillating part of the electron magnetization, $M$, at zero temperature. In general case it has the following form \cite{m-sh19,shen}:
\begin{eqnarray} \label{12}
 M&=&H^{1/2}G\!\left(\!\frac{F}{H}-\phi\right) \nonumber \\
 &\propto&-H^{1/2}\sum_{n=1}^{\infty}\frac{1}{n^{3/2}} \sin\!\left(\!\!2\pi n \!\left[\frac{F}{H}-\phi\right] \pm \frac{\pi}{4}\right),
 \end{eqnarray}
where $G(x)$ is the oscillating function with the period that is equal to unity, $F=S_{\rm ex}c/2\pi\hbar e$ is the frequency determined by the extremal cross-sectional area $S_{\rm ex}$ of the Fermi surface, the phase $\phi$ coincides with the constant $\gamma$,
\begin{eqnarray} \label{13}
 \gamma=\frac{1}{2}-\frac{\Phi_B}{2\pi},
 \end{eqnarray}
that appears in the semiclassical quantization rule specifying the Landau subbands $\varepsilon^l(p_{\parallel})$ in the magnetic field $H$ \cite{Sh,prl},
\begin{equation}\label{14}
S(\varepsilon^{l},p_{\parallel})=\frac{2\pi\hbar e H}{c}\left(l+\gamma\right),
\end{equation}
$\Phi_B$ is the Berry phase of the electron orbit, $p_{\parallel}$ is the quasi-momentum along the magnetic field, and $l$ is a nonnegative integer. The additional offsets $\pm \pi/4$ in Eq.~(\ref{12}) refer to the minimal and maximal $S_{\rm ex}$, respectively, and they result from the expansion of the function $G(x)$ in the Fourier series. The above-mentioned phase shift of the oscillations is due to the Berry phase $\pm \pi$ for the orbits surrounding a band-contact line whereas $\Phi_B=0$ for the orbits which do not link to the line \cite{prl}. It is important that near the crossing point, the orbits in the planes $p_3={\rm const.}>0$ and $p_3={\rm const.}<0$ have the Berry phases of opposite signs, i.e., $\pi$ and $-\pi$. This follows from formulas (19)-(26) of Ref.~\cite{jetp}. Thus, the Berry phase has to change abruptly in the plane $p_3=0$, and one may expect to obtain $\Phi_B=0$ for the orbit in this plane in spite of existence of the nodal line enclosed by the orbit. To investigate the situation in detail, one should  take into account the spin-orbit interaction.

\section{Effect of spin-orbit interaction}

\subsection{Electron spectrum and Fermi surface}

When the weak spin-obit interaction is taken into account, the Hamiltonian of the electron states near the crossing point takes the form \cite{jetp}:
 \begin{eqnarray}\label{15}
\hat H=\left (\begin{array}{cc} \tilde E_{c,c} & \tilde E_{c,v} \\ \tilde E_{c,v}^+ & \tilde E_{v,v}  \\
\end{array} \right),
 \end{eqnarray}
where
 \begin{eqnarray}\label{16}
\tilde E_{c,c}=\left (\varepsilon_{cr}+\Delta_{so}+a p_1+B_2 p_2^2 + B_3 p_3^2 \right )\sigma_0, \nonumber \\
\tilde E_{v,v}=\left (\varepsilon_{cr}-\Delta_{so}+a p_1+B_2 p_2^2 + B_3 p_3^2 \right )\sigma_0, \\
\tilde E_{c,v}=\beta p_2p_3\sigma_0+i(a' p_1+B_2' p_2^2 + B_3' p_3^2 )\sigma_3, \nonumber
\end{eqnarray}
$2\Delta_{so}$ is the gap induced by the spin-orbit interaction in the spectrum at the crossing point ${\bf p}=0$, the energy $\varepsilon_{cr}$ lies in the middle of this gap,  $\sigma_0$ is unit matrix, and $\sigma_3$ is the Pauli matrix.
The energy spectrum corresponding to the Hamiltonian (\ref{15}), (\ref{16}) looks like
\begin{eqnarray}\label{17}
 \varepsilon_{c,v}({\bf p})\!\!&=&\!\varepsilon_{cr}+a p_1+B_2 p_2^2 + B_3 p_3^2+E_{c,v}({\bf p}), \\
 E_{c,v}({\bf p})\!\!&=&\!\pm [\Delta_{so}^2+ (a'p_1+B_2'p_2^2+B_3'p_3^2)^2 \nonumber \\ &+&\beta^2p_2^2p_3^2]^{1/2}. \label{18}
 \end{eqnarray}
These formulas generalize Eqs.~(\ref{1}), (\ref{2}) and differ from them only by the presence of $\Delta_{so}$. As in Sec.~II, we shall set $B_2'=B_3'=a=0$ below. In this case the band-contact lines are the lines along which the energy gap between the electron bands ``$c$'' and ``$v$''  reaches its minimal value $2\Delta_{so}$. With the weak spin-orbit interaction, the Fermi surfaces remain    qualitatively identical to those shown in Figs.~\ref{fig3}-\ref{fig5}. However, at the points of their self-intersection, gaps between the electron and hole pockets of the surfaces appear along the $p_2$ and $p_3$ axes. These gaps $\Delta p_2$ and $\Delta p_3$ are found from the relation $\Delta p_i\approx \Delta_{so}/ |(\zeta - \varepsilon_{cr})B_i|^{1/2}$ where $i=2$, $3$. Due to these gaps, with changing $\zeta$, the transformations of the Fermi surfaces shown in Figs.~\ref{fig3}-\ref{fig5} occur in the interval $\varepsilon_{cr} -\Delta_{so} \le \zeta \le \varepsilon_{cr} +\Delta_{so}$ rather than at the point $\zeta = \varepsilon_{cr}$.

\subsection{Cross-sectional areas}

With nonzero $\Delta_{so}$, formulas (\ref{3}), (\ref{4})  for the cross-sectional areas remain true, but expressions (\ref{5}), (\ref{6}) are modified as follows:
\begin{eqnarray}\label{19}
 x_{a,b}^2&=&\xi_2\xi_{\zeta}-\xi_{\lambda}\tilde p_3^2\left (1-\frac{2}{\lambda}\right ) \nonumber \\
 &\mp& \sqrt{\frac{4\tilde p_3^2}{\lambda}
\left[\xi_3\xi_{\zeta}-\tilde p_3^2\left(1-\frac{1}{\lambda}\right)\right] +\tilde \Delta_{so}^2},
   \end{eqnarray}
where $\tilde p_3 \equiv  |B_3|^{1/2}p_3/ |\zeta-\varepsilon_{cr}|^{1/2}$, $\tilde \Delta_{so}\equiv \Delta_{so}/ |\zeta -\varepsilon_{cr}|$ are the dimensionless quasi-momentum and spin-orbit gap, and the signs minus and plus refer to $x_a$ and $x_b$, respectively. At $\tilde \Delta_{so}\to 0$, formula (\ref{19}) reduces to Eqs.~(\ref{5}), (\ref{6}), and hence at small $\tilde \Delta_{so}$, the function $F_3(\tilde p_3,\tilde \Delta_{so})$ defined by the formulas (\ref{4}), (\ref{19})  practically coincides with $F_3(\tilde p_3)$ calculated in the preceding section.

Consider now the case when $\tilde \Delta_{so}$ is not small ($\tilde \Delta_{so}\lesssim 1$). As in Sec.~II, in the vicinity of the neck ($\tilde p_3 \sim 1$, $\xi_3\xi_{\zeta}=-1$), formulas (\ref{4}), (\ref{19}) can be simplified at small values of  $|\lambda|$:
\begin{eqnarray}\label{20}
 x_{a}^2\!&\approx&\!\frac{|\lambda|}{4\tilde p_3^2}\left [(1+\tilde p_3^2 )^2 -\tilde \Delta_{so}^2 \right ],\ \ \ x_b\approx\frac{2|\tilde p_3|}{|\lambda|^{1/2}},~~~~~~ \nonumber \\
 F_3(\tilde p_3,\tilde \Delta_{so})\!&\approx&\!\frac{3\pi}{4}x_b x_a^2\approx \frac{3\pi|\lambda|^{1/2}[(1+\tilde p_3^2)^2- \tilde \Delta_{so}^2]}{8|\tilde p_3|}.
   \end{eqnarray}
The minimization of the function $F_3(\tilde p_3,\tilde \Delta_{so})$ in Eq.~(\ref{20}) over  $\tilde p_3$ gives the position of the minimal cross section on the neck, $\tilde p_{3,min}$,
 \begin{equation}\label{21}
\tilde  p_{3,min}^{\,2}=-\frac{1}{3}+\sqrt{\frac{4}{9}-\frac{\tilde \Delta_{so}^2}{3}}.
 \end{equation}
The neck exists when $\tilde p_{3,min}^{\,2}>0$. Therefore, in the case of {\it small} $|\lambda|$, the neck occurs at $\tilde \Delta_{so}< 1$, i.e., if $|\zeta-\varepsilon_{cr}|>\Delta_{so}$. Inserting Eq.~(\ref{21}) into Eq.~(\ref{20}), one finds the dependence of the minimal cross-sectional area $S_{3,min}$ on the strength of the spin-orbit interaction.

\begin{figure}[tbp] 
 \centering  \vspace{+9 pt}
\includegraphics[scale=1.0]{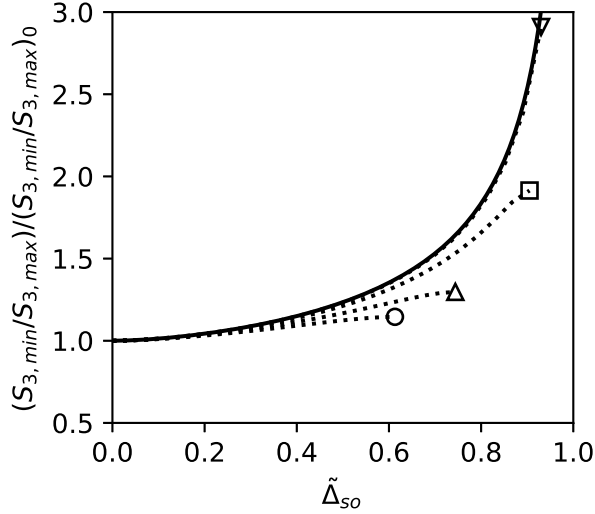}
\caption{\label{fig8} The $\tilde \Delta_{so}$-dependence of the ratio $S_{3,min}/S_{3,max}$ normalized to its value $(S_{3,min}/S_{3,max})_0$ at $\tilde \Delta_{so}=0$. The dotted lines correspond to different negative values of $\lambda$: $-0.01$ ($\nabla$), $-0.1$ ($\Box$), $-0.3$ ($\triangle$), $-0.5$ ($\bigcirc$). The function $f(\tilde \Delta_{so})$ described in the text is shown by the solid line. The $(S_{3,min}/S_{3,max})_0$ is well approximated by Eq.~(\ref{10}).
 } \end{figure}

For $\lambda<0$, apart from $S_{3,min}$, the maximal cross section $S_{3,max}$ exists at $\tilde p_3=0$. Since according to Eq.~(\ref{19}), $x_{a,b}^2=1\mp \tilde \Delta_{so}$ for $\tilde p_3=0$, formula (\ref{4}) gives $F_3(0,\tilde \Delta_{so})$ which differs from its value $F_3(0)=2$ at $\tilde\Delta_{so}=0$,
 \begin{equation} \label{22}
  F_3(0,\tilde \Delta_{so})=2(1+\tilde \Delta_{so})^{1/2}[E(t)-\tilde \Delta_{so}K(t)],
 \end{equation}
where $t=(1-\tilde \Delta_{so})^{1/2}/(1+\tilde \Delta_{so})^{1/2}$, and $E(t)$, $K(t)$ are the complete elliptic integrals. Hence, the ratio $S_{3,min}/S_{3,max}$ for negative $\lambda$ can be represented in the form:
 \begin{equation}\label{23}
  \frac{S_{3,min}}{S_{3,max}}=\frac{F_3(\tilde p_{3,min},\tilde \Delta_{so})}{F_3(0,\tilde \Delta_{so})}= \frac{\pi|\lambda|^{1/2}}{\sqrt{3}}f(\tilde \Delta_{so}),
 \end{equation}
where $\pi|\lambda|^{1/2}/\sqrt{3}$ is the value of  this ratio without spin-orbit interaction (i.e., at $\tilde \Delta_{so}=0$), and the function $f(\tilde \Delta_{so})$ follows from formulas (\ref{20})-(\ref{22}). This function is shown in Fig.~\ref{fig8}. In this figure we also present the $\tilde \Delta_{so}$-dependences of the ratio $S_{3,min}/S_{3,max}$ normalized to its value at $\tilde \Delta_{so}=0$ for not-too-small values of $|\lambda|$. Note that the appropriate curves terminate at certain values of $\tilde \Delta_{so}<1$ since at these critical $\tilde \Delta_c$, the cross sections with areas $S_{3,min}$ and $S_{3,max}$ merge, i.e., $\tilde p_{3,min}$ reaches zero and $S_{3,min}=S_{3,max}$. The values of $\tilde \Delta_c(\lambda)$ are determined from the relationship:
\begin{equation}\label{24}
|\lambda| = \frac{2[K(t)-E(t)](1+ \tilde{\Delta}_c)}{(1+ \tilde{\Delta}_c)E(t)-\tilde{\Delta}_c K(t)},
\end{equation}
where $t = \sqrt{(1-\tilde{\Delta}_c)/(1+ \tilde{\Delta}_c)}$. For small $|\lambda|$, we obtain from this relationship that $\tilde \Delta_c \approx 1-|\lambda|$, and $\tilde \Delta_c$ decreases with increasing $|\lambda|$. When $\tilde \Delta_c <\tilde \Delta_{so}<1$, i.e., when the chemical potential lies in the interval: $\Delta_{so}<|\zeta-\varepsilon_{cr}| <\Delta_{so}/\tilde \Delta_c$, the extremal cross section exists only at $\tilde p_3=0$, and the area of this cross section is still described by Eq.~(\ref{22}). However, this area is minimal in $p_3$ now.

\subsection{Cyclotron masses}

\begin{figure}[tbp] 
 \centering  \vspace{+9 pt}
\includegraphics[scale=1.0]{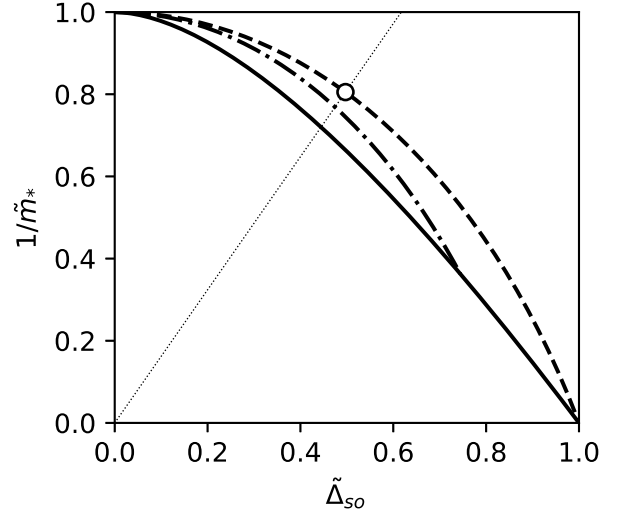}
\caption{\label{fig9} The $\tilde\Delta_{so}$-dependences of the  inverse dimensionless masses $[\tilde m_{*,min}]^{-1}$ and $[\tilde m_{*,max}]^{-1}$. The $[\tilde m_{*,min}(\tilde\Delta_{so})]^{-1}$ at $|\lambda|\ll 1$ (the dashed line) and $[\tilde m_{*,max}(\tilde\Delta_{so})]^{-1}$ (the solid line) are defined by Eqs.~(\ref{25}) and (\ref{26}), respectively. These equations at small $\tilde\Delta_{so}$ give $[\tilde m_{*,min}(\tilde\Delta_{so})]^{-1} \approx 1-0.75\tilde\Delta_{so}^2$ and  $[\tilde m_{*,max}(\tilde\Delta_{so})]^{-1}\approx 1-\tilde\Delta_{so}^2\ln(4/\sqrt{2\tilde\Delta_{so}})$.
The dash-and-dot line depicts $[\tilde m_{*,min}(\tilde\Delta_{so})]^{-1}$ for $\lambda=-0.3$.
The dotted line shows the left hand side of Eq.~(\ref{37}) for  $\kappa=1.62$, and the circle marks the solution of this equation.
 } \end{figure}

Consider now the cyclotron masses corresponding to the orbits with the areas $S_{3,min}$ and $S_{3,max}$. In the case of $S_{3,min}$ we obtain for {\it small} $|\lambda|$,
\begin{eqnarray}\label{25}
 |m_{*,min}|&=&\frac{3S_{3,min}}{4\pi |\zeta-\varepsilon_{cr}|}\cdot \tilde m_{*,min}, \nonumber \\
 \tilde m_{*,min}&=&\frac{(1+\tilde p_{3,min}^{\,2})^2 +\frac{1}{3}\tilde \Delta_{so}^2}{(1+\tilde p_{3,min}^{\,2})^2 -\tilde \Delta_{so}^2},
  \end{eqnarray}
where $\tilde p_{3,min}^{\,2}$ is given by Eq.~(\ref{21}).
For not-too-small $\lambda$, the function $\tilde m_{*,min}(\tilde\Delta_{so})$ has a complicated form, and it can be calculated with Eqs.~(\ref{4}) and (\ref{19}).
In the case of $S_{3,max}$ we arrive at
\begin{eqnarray}\label{26}
 |m_{*,max}|\!&=&\!\frac{3S_{3,max}}{4\pi |\zeta-\varepsilon_{cr}|}\cdot \tilde m_{*,max}, \nonumber \\ \tilde m_{*,max}&=&\frac{[E(t)(1+\tilde \Delta_{so})- \tilde \Delta_{so}K(t)]}{[E(t)- \tilde\Delta_{so} K(t)](1+\tilde \Delta_{so})},
  \end{eqnarray}
where $t=\sqrt{(1-\tilde \Delta_{so})/(1+\tilde \Delta_{so})}$. The $\tilde \Delta_{so}$-dependences of $1/\tilde m_{*,min}$ and $1/\tilde m_{*,max}$ are shown in Fig.~\ref{fig9}. At given value of the spin-orbit gap $2\Delta_{so}$, these dependences enable one to determine $|\zeta-\varepsilon_{cr}|$ if the appropriate area and the cyclotron mass are known from an experiment, see the next section.

\subsection{Phase of quantum oscillations}

We now discuss the phase of the oscillations. With the  spin-orbit interaction, the quantization rule (\ref{14}) transforms into the form \cite{Sh}:
\begin{equation}\label{27}
S(\varepsilon^{l},p_{\parallel})=\frac{2\pi\hbar e H}{c}\left(l+\frac{1}{2}\pm \frac{gm_*}{4m}\right),
\end{equation}
where $g$ is the electron $g$ factor; $m$ and $m_*$ are the electron and cyclotron masses, respectively, and the other quantities are defined as in formula (\ref{14}). The phases $\phi$, Eq.~(\ref{12}), for the electrons with the oppositely directed spins are now given by $\phi= (1/2)\pm \delta$ where $\delta \equiv (gm_*/4m)$.
It is clear that the {\it semiclassical} spectrum specifying by Eq.~(\ref{27}) and the phase $\phi$ of the oscillations are, in fact, determined by the fractional part of $\delta$ since any integer can be added to $\delta$ and $\phi$. Moreover, the spectrum and the phase remain unchanged under the replacement $\delta$ by $1-\delta$, and this further reduces the range of nonequivalent values of $\delta$.
It is also worth noting that the contribution of $(gm_*/4m)\equiv \delta$ into the harmonics of the oscillating magnetization $M$, Eq.~(\ref{12}), can be rewritten as the well-known spin factor $\cos(2\pi n \delta)$ \cite{Sh},
\begin{eqnarray*}
   M \propto -H^{1/2}\sum_{n=1}^{\infty}\frac{\cos(2\pi n \delta)}{n^{3/2}}  \sin\!\left(\!\!2\pi n \!\left[\frac{F}{H}-\frac{1}{2}\right] \pm \frac{\pi}{4}\right).
 \end{eqnarray*}

The theory of the $g$ factor for itinerant electrons was elaborated in Refs.~\cite{jetp,g1}. At the  weak spin-orbit interaction, the $g$ factor comprises the two terms,  $g=g_1+g_2$. The first term $g_1$ is large for the orbits surrounding a band-contact line and is determined by their Berry phase, while the second one, $g_2$, is specified by an interband part $L$ of the electron orbital moment (if one considers a semiclassical electron as a wave packet, this $L$ can be interpreted as the orbital moment associated with self-rotation of the wave packet around its center of mass). Neglecting the Zeeman term describing the direct interaction of the electron spin ${\bf s}$ with the magnetic field, $e\hbar{\bf s}\cdot{\bf H}/mc$, it was shown \cite{jetp,g1} that the {\it total} $g$ factor, $g=g_1+g_2$, has the universal value, $g=2m/m_*$, for any electron orbit surrounding an isolated band-contact line in a crystal with the weak spin-orbit interaction. Insertion of this universal value into formula (\ref{27}) reproduces equation (\ref{14}) with $\gamma=0$, and leads to $\phi=0$. This means that when the weak  spin-orbit interaction is `` turned on'' in a crystal, the quantum oscillations can gradually vary in their frequency and magnitude, but their phase, which serves as the topological characteristic of the band-contact lines in absence of the interaction, remains unchanged.
It is this result that justifies the use of the concept of the band-contact line in presence of the weak spin-orbit interaction which generally lifts the accidental  contact of the bands along the lines.

Let us now calculate the $g$ factor for the orbits near the crossing point, i.e., in the region where the dispersion law given by Eqs.~(\ref{17}), (\ref{18}) noticeably  differs from the Dirac spectrum. When the spin-orbit interaction is weak, formula (19) of Ref.~\cite{g1} is applicable for the calculation of $\delta\equiv gm_{*}/4m$,
 \begin{eqnarray}\label{28}
\delta=-\frac{1}{2\pi}\oint_\Gamma \frac{d{\bf p}_{\bot} }{v_{\bot }}\mu_{0,11}({\bf p}) = -\frac{1}{2\pi}\oint_\Gamma \frac{dp_2 }{|v_1|}\mu_{0,11}({\bf p}),
 \end{eqnarray}
where $v_\bot $ is the absolute value of projection of electron velocity ${\bf v}$ on the plane of the orbit $\Gamma$ defined here by the condition $p_3=$const., $d{\bf p}_{\bot}$ is the infinitesimal element of the orbit,
\begin{eqnarray}\label{29}
\mu_{0,11}\!\!&=& -\frac{v_1\beta p_3}{2a'p_1} \frac{(\zeta -\varepsilon_{cr}+\Delta_{so}+B_2p_2^2-B_3p_3^2)}{(\zeta -\varepsilon_{cr}+\Delta_{so}-B_2p_2^2-B_3p_3^2)}, \\
 v_1\!\!&=&\frac{a'^{\,2}p_1}{(\zeta -\varepsilon_{cr} -B_2p_2^2-B_3p_3^2)}. \label{30}
 \end{eqnarray}
Here the component $v_1$ of the velocity has been found with Eqs.~(\ref{17}), (\ref{18}) at $a=0$, and the quantity $\mu_{0,11}$  has been calculated with the formulas of the Appendix of Ref.~\cite{g1}. The orbit $\Gamma$, i.e., the function $p_1(p_2)$ is obtained from the equation:
 \begin{eqnarray}\label{31}
 \tilde p_1^2=(x_a^2-\tilde p_2^2)(x_b^2-\tilde p_2^2),
 \end{eqnarray}
where $\tilde p_1 \equiv a'p_1/|\zeta-\varepsilon_{cr}|$, $\tilde p_{2,3} \equiv  |B_{2,3}|^{1/2}p_{2,3}/ |\zeta -\varepsilon_{cr}|^{1/2}$ are the dimensionless quasi-momenta, $\tilde p_2^2 \le x_a^2$, and  $x_a^2$, $x_b^2$ are determined by Eq.~(\ref{19}). Inserting formulas (\ref{29})-(\ref{31}) into Eq.~(\ref{28}), we eventually arrive at
 \begin{eqnarray}\label{32}
\delta=\pm \frac{2\tilde p_3}{\pi|\lambda|^{1/2}x_b} \left ( -K(t)+ 2\Pi(\nu;t)\right ),
\end{eqnarray}
where the signs plus and minus refer to the electron and hole orbits, respectively, $t=x_a/x_b$,
 \[
 \nu=\xi_2\xi_{\zeta}\frac{(1-\xi_{\zeta}\tilde \Delta_{so} -\xi_3\xi_{\zeta}\tilde p_3^2)}{x_b^2},
  \]
$\xi_2$, $\xi_3$, $\xi_{\zeta}$ are the signs of $B_2$, $B_3$, $\zeta-\varepsilon_{cr}$, and
\begin{eqnarray*}
K(t)&=&\int_0^1\frac{du}{(1-u^2)^{1/2}(1-t^2u^2)^{1/2}},\\
\Pi(\nu;t)&=&\int_0^1\frac{du}{(1-u^2)^{1/2}(1-t^2u^2)^{1/2}(1-\nu u^2)}
\end{eqnarray*}
are the complete elliptic integrals of the first and the third kinds. Below we shall be mainly interested in $\{\delta\}$, the fractional part of $\delta$, since this part actually determines the phase of the quantum oscillations.

\begin{figure}[tbp] 
 \centering  \vspace{+9 pt}
\includegraphics[scale=1.0]{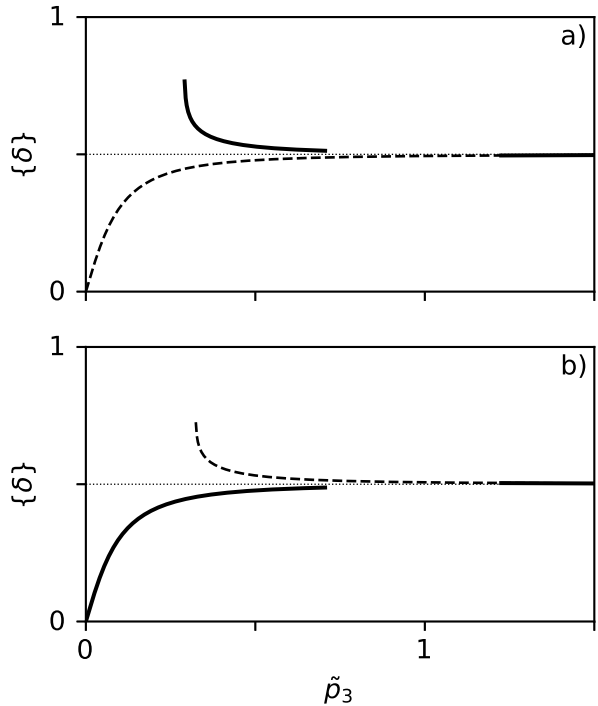}
\caption{\label{fig10} Dependences of $\{\delta\}$, the fractional part of $\delta$, on the coordinate $\tilde p_3$ of the orbit at $\lambda=-0.1$, $B_2<0$, $B_3>0$ (a) and at $\lambda=0.1$, $B_2>0$, $B_3>0$ (b); $\tilde \Delta_{so}=0.5$, the magnetic field is directed along the $p_3$ axis. The solid and dashed lines correspond to the cases  $\zeta-\varepsilon_{cr}>0$ and $\zeta-\varepsilon_{cr}<0$, respectively. The dotted line shows the case $\tilde\Delta_{so}=0$. Values of $\{ \delta\}$ at $B_3<0$ can be found, using the invariance of the fractional part of $\delta$ under the transformation:
$B_{2,3}\to -B_{2,3}$, $\zeta-\varepsilon_{cr}\to -(\zeta-\varepsilon_{cr})$.
} \end{figure}

Dependences of $\{ \delta\}$ on the coordinate $\tilde p_3$ of the orbit  at $\tilde\Delta_{so}=0.5$ and $\lambda=\pm 0.1$ are shown in Fig.~\ref{fig10}.  As was supposed in the preceding section, the $\tilde p_3$-dependence of $\{ \delta\}$ does pass through zero at $\tilde p_3=0$. However, our analysis shows that the width of the $\tilde p_3$-region where $\{ \delta\}$ deviates from  $1/2$  decreases with decreasing $|\lambda|$ and $\tilde\Delta_{so}$. The termination of the lines in Fig.~\ref{fig10} at the finite nonzero $\tilde p_3$ is caused by the absence of the closed orbits at small $\tilde p_3$ in the  cases $\lambda<0$, $(\zeta-\varepsilon_{cr})B_3>0$ and $\lambda>0$, $(\zeta-\varepsilon_{cr})B_3<0$; see Figs.~\ref{fig4} and \ref{fig5}. On the other hand, the interruption of the solid lines is due to the disappearance of the self-intersection and appearance of the gap between the electron and hole pockets of the Fermi surface at nonzero spin-orbit interaction.

For the {\it extremal} orbits lying on the necks of the Fermi surface, the quantity $\delta$ is the function of $\tilde \Delta_{so}$ and $\lambda$ only. For such orbits, in Fig.~\ref{fig11}, we show the dependences of $\{ \delta\}$ on $\tilde \Delta_{so}$ at various negative values of $\lambda$ and the dependence of $\{\delta\}$ on $\lambda$ at fixed $\tilde \Delta_{so}$. It is seen that at small $|\lambda|$, the values of $\{\delta\}$ are close to $1/2$ practically for all $\tilde\Delta_{so}$, and in this case formula ({\ref{32}) for $\tilde p_3>0$ yields
  \begin{eqnarray}\label{33}
 \delta\!&\approx&\!\pm \left(\frac{1}{2} -\xi_3 \frac{\lambda \tilde \Delta_{so}}{8\tilde p_{3,min}^{\,2}}\right), \nonumber\\
 \{ \delta\}\!&\approx&\!\frac{1}{2} + \frac{\lambda \tilde \Delta_{so}}{8\tilde p_{3,min}^{\,2}},
\end{eqnarray}
where $\tilde p_{3,min}^{\,2}$ is given by Eq.~(\ref{21}), and we have taken into account that $\xi_3=-1$ for the electron Fermi-surface neck and $\xi_3=1$ for the hole one.
The essential deviation of $\{ \delta\}$ from $1/2$ occurs only when $\tilde \Delta_{so}$ tends to  its critical value $\tilde \Delta_c$ determined by Eq.~(\ref{24}), i.e., when the neck approaches the crossing point.

\begin{figure}[tbp] 
 \centering  \vspace{+9 pt}
\includegraphics[scale=1.0]{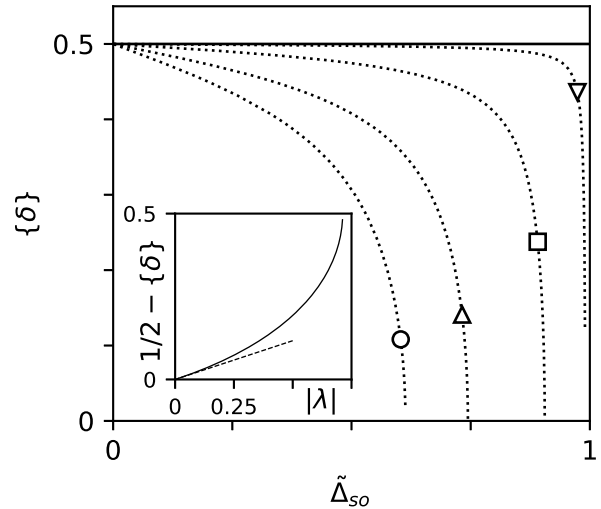}
\caption{\label{fig11} The $\tilde \Delta_{so}$-dependences of $\{\delta\}$ (the dotted lines)  for the extremal orbits lying on the Fermi-surface necks at $\lambda= -0.01$ ($\nabla$), $-0.1$ ($\Box$), $-0.3$ ($\triangle$), $-0.5$ ($\bigcirc$). The value of $\{\delta\}$ reaches zero at $\tilde \Delta_{so}=\tilde \Delta_c$ described by Eq.~(\ref{24}). The inset: The solid line shows the dependence of $(1/2)-\{\delta\}$ on $|\lambda|$ at $\tilde \Delta_{so}=0.5$, the dashed line corresponds to Eq.~(\ref{33}) with the same $\tilde \Delta_{so}$.
} \end{figure}

When $\{\delta\}$ tends to zero or $1/2$, the phases $\phi=1/2 + \delta$ and $\phi=1/2- \delta$ of the quantum oscillations coincide with each other and are equal to $1/2$ or $0$, respectively. Therefore, to distinguish between the orbits on the necks of the Fermi surface ($\{\delta\} \approx 1/2$, $\phi\approx 0$) and the orbits lying in the planes passing through the crossing point ($\{\delta\}=0$, $\phi=1/2$), it is sufficient to measure the first harmonic of the oscillations and to find its phase $\phi$. However, when the magnetic fields are sufficiently strong, many harmonics contribute to the oscillations which are now not sinusoidal, and in this situation the phase $\phi$ is usually found with the Landau-level fan diagram \cite{Sh}. In particular, the oscillating magnetization at strong magnetic fields  exhibits sharp minima or sharp maxima when the edges of the Landau subbands cross the Fermi level, i.e., when $(F/H)-\phi$ is equal to integer numbers \cite{shen}. The sharp minima in the oscillations are produced by a minimal cross section whereas the sharp maxima correspond to a maximal cross section, see, e.g.,  Fig.~1 in Ref.~\cite{prl04}. Thus, not only $\phi$ but also this feature of the oscillating magnetization can be useful in distinguishing between the two types of the orbits. However, it should be noted that the Landau-level fan diagram can lead to a discrepancy between the measured phase $\phi$ and its theoretical value if this  diagram is plotted in a magnetic-field range in which the shape of the oscillations changes. This is clear from the following considerations: At low magnetic fields when the first harmonic gives the main contribution to the oscillations, the Landau-level fan diagram plotted with the minima of the magnetization is described by the equation resulting from formula (\ref{12}):
\begin{eqnarray*}
2\pi\left(\frac{F}{H_n}-\phi+\frac{1}{8}\right)=2\pi(n+\frac{1}{4}),
\end{eqnarray*}
where $n$ is the Landau-level index, the fields $H_n$ correspond to the minima of the magnetization, and for definiteness, we have used the offset $\pi/4$ characteristic of the orbits on the Fermi-surface necks. In this equation the minima correspond to $n+1/4$, and the extrapolation of the linear dependence of $F/H_n$ on $n$ to the value $F/H_n = 0$ leads to the following intercept $n_0$ in the $n$ axis:
\begin{eqnarray*}
 n_0=-\frac{1}{8}-\phi.
\end{eqnarray*}
On the other hand, at strong magnetic fields when many harmonics contribute to the oscillating magnetization, and its shape differs from the sinusoid, the oscillations exhibit sharp minima at integer $n$, and one has
 \begin{eqnarray*}
\frac{F}{H_n}-\phi=n,
\end{eqnarray*}
i.e, the Landau-level fan diagram plotted with $H_n$ from the strong-field region gives the intercept, $n_0=-\phi$. In other words, the change in the shape of the oscillations gradually varies the correspondence between the Landau index $n$ and the minima in the magnetization, and this variation can lead to a spurious value of $\phi$ found with the diagram. Note that in Ref.~\cite{Hu1}, the change in the intercept $n_0$ was really observed when the magnetic-field range of the Landau-level fan diagram was extended.

If the fractional part of $\delta$ noticeably differ from zero and $1/2$, the {\it sharp} minima  (maxima) of the oscillations tend to split at strong magnetic fields. In this case, it is preferable to measure the $g$ factor rather than $\phi$.  A measurement of the $g$ factor (the quantity $\delta$) requires either an analysis of both the first and the second harmonics of the oscillations or a direct detection of the quantum-oscillation splitting \cite{Sh}. In the latter case, this splitting $(F/H_+)-(F/H_-)$ is equal to $2\delta$ or $1-2\delta$  where the fields $H_+$ and $H_-$ mark the split sharp minima (maxima). Measurements of $\delta$ can, in principle, provide addition information on the parameters of the spectrum $\tilde\Delta_{so}$ and $\lambda$, see, e.g., formula (\ref{33}).

\section{Example: ${\rm ZrSiS}$}

We now apply the obtained results to ZrSiS. According to the  band-structure calculations \cite{pez}, the following sequence of the crossing-points energies occurs in this semimetal: $\varepsilon_{cr}^{U} > \varepsilon_{cr}^{\Delta}>\zeta > \varepsilon_{cr}^{\Sigma} > \varepsilon_{cr}^{S}$, with the energies $\varepsilon_{cr}^{\Delta}$, $\varepsilon_{cr}^{\Sigma}$, and the chemical potential $\zeta$ being very close to one another. The Fermi surface consists of the hole and electron ``tubes'' enclosing  the nodal lines U-$\Delta$ and $\Sigma$-S, respectively, and of the self-intersecting parts penetrated by the lines $\Delta$-$\Sigma$ and U-S, Fig.~\ref{fig1}. The results of Secs.~II and III can be useful in analyzing an extremal orbit if it is close to one of the crossing points and is small as compared to the distances  between this point and the neighboring ones. For ZrSiS, this condition is fulfilled for the orbits on the tubes' necks located near  the points $\Sigma$ and $\Delta$. The Fermi surface near these points refers to the type shown in Fig.~\ref{fig5} ($\lambda<0$), and these orbits can reveal themselves in the quantum oscillation of various physical quantities  when the direction of the magnetic field is close to $\Gamma$-Z axis (i.e., to the c-axis of the crystal). As to the point S, in the vicinity of which the Fermi surface refers to the type presented in Fig.~\ref{fig3} ($\lambda>0$, $\zeta-\varepsilon_{cr}^S>0$) \cite{com2}, there is no neck near this point, whereas the extremal orbits in the reflection planes of the crystal are large, and hence the results of Secs.~II and III are not appropriate for calculating their characteristics. It should be also noted that Eqs.~(\ref{1}), (\ref{2}) hardly describes the spectrum near the crossing point $U$  since near the crossing energy $\varepsilon_{cr}^U$, one more energy band exists in ZrSiS  \cite{schoop,pez,fu19}.

\begin{table}
\caption{The frequencies $F_i$, the cyclotron masses $m_{*,i}$, and phases $\phi_i$ ($i=1, 2$) of the quantum oscillations measured in ZrSiS at $H\parallel$ c. The values of $\phi_i$ have been found from the experimental data of Refs.~\cite{wang1,singha,Hu1,matus}, using the definition of the phase $\phi$ as in Eq.~(\ref{12}). The quantities $|\zeta-\varepsilon_{cr}|_{i,0}$ and $m^2(a')^2|B_2| /|\lambda|$  are calculated with Eqs.~(\ref{11}) and (\ref{36}), respectively, neglecting the spin-orbit coupling.}
\begin{tabular}{l|ccccc}
Reference&\cite{Ali1}&\cite{wang1}&\cite{singha}&\cite{Hu1} &\cite{matus}\\ \colrule
$F_1$(T)&-&-&-&8.4&8.5\\
$m_{*,1}/m$&-&-&-&0.025&0.07 \\
$\phi_1$&-&-&-&$0.53$\footnote{This value corresponds to the offset $\pi/4$ in Eq.~(\ref{12}), see the text.}&$-\frac{1}{4}$ or $\frac{1}{4}$\\
$|\zeta-\varepsilon_{cr}|_{1,0}$\ (meV)
&-&-&-&58&21 \\
$\frac{m^2(a')^2|B_2|}{|\lambda|}$\ (eV)&-&-&-&122&5.7 \\
 \colrule
$F_2$(T) &23&18.9&14&-&15.3 \\
$m_{*,2}/m$&0.11&0.12&0.1&-&0.14 \\
$\phi_2$&-&$0\pm 0.14$&$0\pm 0.2$&-&-\\
$|\zeta-\varepsilon_{cr}|_{2,0}$\ (meV)&36&27&24&-&19 \\
$\frac{m^2(a')^2|B_2|}{|\lambda|}$\ (eV)&3.9&2.4&3.1&-&1.3\\
\end{tabular}
\end{table}

The cross-sectional areas corresponding to low-frequency quantum  oscillations, the appropriate cyclotron masses, and the phases of the oscillations were measured  in ZrSiS for the magnetic fields parallel to the c-axis \cite{Hu1,Ali1,wang1,singha,matus}, Table I.
Let us now analyze various possible interpretations of these experimental data, pointing out evidences for and against each interpretation.

\subsection{Extremal cross sections are on two different necks}

The frequency $F_1\approx 8.5$T observed in the de Haas - van Alphen \cite{Hu1} and thermoelectric-power \cite{matus} oscillations appears to correspond to the minimal cross-sectional area $S_{min}$ associated with one of the two above-mentioned necks. This hypothesis is supported by the following experimental results: 1) If the magnetic field is tilted away from the \mbox{c-axis} at a little angle $\theta$, one has $S_{min}(\theta)\approx S_{min}(0)/\cos\theta$ \cite{Hu1}; cf. Sec.~II.
2) At low temperatures and high magnetic fields, when many harmonics contribute to the oscillations, the magnetization reveals sharp minima, cf. Fig.~1b,c in Ref.~\cite{Hu1} and Fig.1 in Ref.~\cite{prl04}. As was mentioned in the preceding section, this feature of the magnetization is just inherent in the case of a neck on the Fermi surface \cite{shen}. The value of the phase $\phi_1$ testifying against this hypothesis will be discussed below. Another frequency $F_2\sim 14\div23$ T was detected in the oscillations of the resistivity \cite{Ali1,wang1,singha} and of the thermoelectric power \cite{matus}. It is important that in Ref.~\cite{matus}, the frequencies $F_1$ and $F_2$ were found in one and the same sample. This means that the difference between $F_1$ and $F_2$ cannot be ascribed to dissimilar doping in the different samples, and hence, these frequencies correspond to different cross sections. The values of $\phi_2$, see Table I, and absence of a noticeable angular dependence of $F_2$ at angles $\theta \lesssim 20^{\circ}$ \cite{singha} indicate that the frequency $F_2$ seems to correspond to the second neck mentioned above.

Using formula (\ref{11}), in which $S_3(\zeta,p_{3,ex})=2\pi\hbar eF_i/c$,  and the cyclotron masses $m_{*,i}$, we calculate the position of the chemical potential relative to the energy of the  crossing point, $|\zeta-\varepsilon_{cr}|_{i,0}$, see Table I. The subscript zero means that this quantity is calculated without considering the spin-orbit interaction. The large difference in $|\zeta-\varepsilon_{cr}|_{1,0}$ for the data of Refs.~\cite{Hu1} and \cite{matus} is due to the essential difference in $m_*$ at practically coinciding frequencies in these papers. This discrepancy in the cyclotron masses, which were obtained from the oscillations of the magnetization \cite{Hu1} and of the thermoelectric power \cite{matus}, cannot be explained by dissimilar doping in the samples and requires an additional experimental investigation.

According to the last column in Table I, the difference $\varepsilon_{cr}^{\Delta} -\varepsilon_{cr}^{\Sigma} \approx 40$ meV, and the chemical potential lies practically in the middle of the interval from $\varepsilon_{cr}^{\Sigma}$ to $\varepsilon_{cr}^{\Delta}$. Using this $\varepsilon_{cr}^{\Delta} -\varepsilon_{cr}^{\Sigma}$, and also  $\varepsilon_{cr}^U-\varepsilon_{cr}^{\Delta}\sim 0.3$ eV, $\varepsilon_{cr}^{\Sigma}-\varepsilon_{cr}^{S}\sim 0.1$ eV derived from Fig.~1 in Ref.~\cite{pez}, one can roughly estimate $B_2$, $B_3$, and $\lambda$ for the points $\Sigma$ and $\Delta$, assuming that $B_n^i p_{ij}^2\sim \varepsilon_{cr}^j- \varepsilon_{cr}^i$ where $j$ is the adjacent (along the axis $n=2$ or $3$) crossing point to the point $i$, and $p_{ij}$ is the distance between these points in the Brillouin zone. We take $(1/\sqrt{2})\pi\hbar/a$ as the distance between the points $\Sigma$ and $\Delta$, and $\pi\hbar/c$ as the distances U-$\Delta$ and $\Sigma$-S where $c=8.07$\,\AA, $a=3.55$\,\AA\ are the crystal-lattice parameters of the ZrSiS \cite{wang1}. Then, we arrive at the estimates: $-B_2^{\Delta}\sim B_2^{\Sigma}\sim 0.015/m$, $B_3^{\Delta}\sim 0.3/m$, $B_3^{\Sigma}\sim -0.1/m$ where $m$ is the electron mass \cite{com}. Hence, if $\beta\sim 1/m$, we obtain $\lambda^{\Delta}\sim -0.018$, $\lambda^{\Sigma}\sim -0.006$. A somewhat more accurate estimates give \cite{com3}:
  \begin{eqnarray} \label{34}
 -B_2^{\Delta}&\sim& B_2^{\Sigma}\sim \frac{0.037}{m},\ \ B_3^{\Delta}\sim \frac{0.74}{m}, \ \ B_3^{\Sigma}\sim -\frac{0.25}{m}, \nonumber \\
 \lambda^{\Delta}&\sim& -\frac{0.11}{(m\beta_{\Delta})^2}, \ \ \lambda^{\Sigma}\sim -\frac{0.037}{(m\beta_{\Sigma})^2}, \\
  a'_{\Sigma}&\sim&  7.7\times 10^5 \frac{\rm m}{\rm s},\ \ a'_{\Delta}\sim 4.3\times 10^5 \frac{\rm m}{\rm s}, \nonumber
\end{eqnarray}
where for completeness we have also presented the values of $a'_{\Sigma}$ and $a'_{\Delta}$ that follow from the electron-band structure shown in Fig.~1 of Ref.~\cite{pez}. Of course, reliable values of all these  parameters can be obtained by a fitting of spectrum (\ref{17}), (\ref{19}) to results of the band structure calculations near the point $\Sigma$ and $\Delta$. However, the small magnitudes of $\lambda^i$ found here are actually due to that $\varepsilon_0 \sim p_{ij}^2/m \gg |\varepsilon_{cr}^j- \varepsilon_{cr}^i|= \varepsilon_{max} -\varepsilon_{min}$, and hence this smallness of $|\lambda|$ is the typical feature of the nodal-line semimetals.

Assuming that the parameter $|\lambda|$ is sufficiently small, one can find the value of the following combination of the parameters:
\begin{eqnarray}\label{35}
\frac{m^2(a')^2|B_2|}{|\lambda|}=\frac{(a')^2m^2\beta^2}{4|B_3|},
\end{eqnarray}
using Eqs.~(\ref{3}) and (\ref{9}) rewritten in the form:
\begin{eqnarray}\label{36}
 \mu_BF&=&\frac{|\zeta-\varepsilon_{cr}|_0^{3/2}}{3\pi m a'|B_2|^{1/2}} F_3(\tilde p_{3,min}) \nonumber \\
 &=&\frac{2}{3\sqrt{3}}\frac{|\lambda|^{1/2}|\zeta- \varepsilon_{cr}|_{0}^{3/2}}{ m a'|B_2|^{1/2}} ,~~~
\end{eqnarray}
where $\mu_B$ is the Bohr magneton, $F=cS_{3,min}/(2\pi e\hbar)$ is the frequency of the oscillations, $m$ is the electron mass, and the appropriate difference $|\zeta-\varepsilon_{cr}|_{0}$ is given in Table I. On the other hand, one can estimate this combination for the points $\Sigma$ and $\Delta$, using Eqs.~(\ref{34}),
\begin{eqnarray*}
\frac{m^2(a'_{\Sigma})^2|B_2^{\Sigma}|}{|\lambda^{\Sigma}|}&\sim& 3.3(m\beta_{\Sigma})^2 \ \ \ ({\rm eV}), \\ \frac{m^2(a'_{\Delta})^2|B_2^{\Delta}|}{|\lambda^{\Delta}|}&\sim& 0.35(m\beta_{\Delta})^2 \ \ ({\rm eV}).
\end{eqnarray*}
Comparing these estimates with the data of Table I, we may conjuncture that the frequency $F_1$  corresponds to the extremal orbit on the neck located near the point $\Sigma$, whereas the frequency $F_2$ is produced by the orbit on the neck near the crossing point $\Delta$. In this case we obtain the reasonable values of the parameter $\beta$:
 \[
  m\beta_{\Sigma}\sim 1.3,\ \ \ \ m\beta_{\Delta}\sim 1.9\div 3.3
  \]
(the value $m^2(a'_{\Sigma})^2|B_2^{\Sigma}| /|\lambda^{\Sigma}|\sim 122$ eV seems too large, and we exclude it from the consideration). With these $\beta$, we eventually arrive at the estimates:
 \[
\lambda^{\Sigma} \sim -0.022, \ \ \ \  \lambda^{\Delta} \sim -(0.01\div 0.03),
 \]
which justify the assumption of small $\lambda$.

Recently, the spin-orbit gap in ZrSiS, $2\Delta_{so}=26$ meV, was measured with the magneto-optical spectroscopy \cite{uykur}. Although this value gives the gap averaged along the lines S--U and $\Sigma$--$\Delta$,  we shall use it below in the estimates near the points $\Sigma$ and $\Delta$. To take into account the effect of the spin-orbit coupling on the extremal orbit associated with a frequency $F$, we solve the following equation in $\tilde\Delta_{so}\equiv \Delta_{so}/|\zeta-\varepsilon_{cr}|$:
 \begin{eqnarray}\label{37}
\kappa \tilde\Delta_{so}=\frac{1}{\tilde m_{*,min}(\tilde\Delta_{so})},
 \end{eqnarray}
where $\zeta-\varepsilon_{cr}$ is chemical potential measured from  the middle of the spin-orbit gap, and $\kappa$ is the constant,
 \[
\kappa=\frac{3S_{3,min}}{4\pi m_{*,min}\Delta_{so}}=\frac{3e\hbar F}{2cm_{*,min}\Delta_{so}}=\frac{|\zeta- \varepsilon_{cr}|_{0}}{\Delta_{so}}.
 \]
Equation (\ref{37}) immediately follows from formulas (\ref{25}), and Fig.~\ref{fig9} shows a graphic solution of this equation.
The solutions of Eq.~(\ref{37}) for different $|\zeta- \varepsilon_{cr}|_{i,0}$ from Table I are presented in Table II. Knowing $\tilde\Delta_{so}$, we find the position of the chemical potential relative to the middle of the gap at the crossing point, $|\zeta- \varepsilon_{cr}|_{i}=\Delta_{so}/\tilde\Delta_{so}$ (and hence the position of $\zeta$ relative to the edge of the electron or hole band at this point, $|\zeta- \varepsilon_{cr}|_{i}- \Delta_{so}$). The coordinate $\tilde p_{3,min}$ of the extremal orbit on the neck of the Fermi surface is calculated with Eq.~(\ref{21}), and the value of the combination (\ref{35}) follows from the equation that generalizes formula (\ref{36}) to the case of the nonzero spin-orbit interaction,
\begin{eqnarray}\label{38}
 \mu_BF&=&\frac{|\zeta-\varepsilon_{cr}|^{3/2}}{3\pi m a'|B_2|^{1/2}} F_3(\tilde p_{3,min},\tilde\Delta_{so}) \nonumber \\ &=&\frac{|\lambda|^{1/2}|\zeta-\varepsilon_{cr}|^{3/2}}{ m a'|B_2|^{1/2}} \frac{[(1+\tilde p_{3,min}^2)^2-\tilde\Delta_{so}^2]}{8|\tilde p_{3,min}|},~~~
\end{eqnarray}
where we have used Eq.~(\ref{20}). Similarly to the case without  the spin-orbit coupling, we obtain the values of the parameters $\beta$ and $\lambda$:
\begin{eqnarray*}
  m\beta_{\Sigma}&\sim& 1.5,\ \ \ \ \ \ \ \ \  m\beta_{\Delta}\sim 2.3\div 3.7, \\ \lambda^{\Sigma} &\sim& -0.016, \ \ \ \   \lambda^{\Delta} \sim -(0.008\div 0.021).
 \end{eqnarray*}
It is seen that the analysis of the experimental data with consideration for the spin-orbit interaction leads to the modification of the values of the parameters and of $|\zeta-\varepsilon_{cr}|$. However, this modification  does not change the order of the magnitude of these quantities, even though  $\Delta_{so} \sim |\zeta-\varepsilon_{cr}|_0$. If $\Delta_{so}$ were essentially less than $|\zeta-\varepsilon_{cr}|_0$, the modification would be small.

\begin{table}
\caption{Solutions of Eq.~(\ref{37}) for the values of $|\zeta- \varepsilon_{cr}|_{i,0}$ from Table I. The quantity  $|\zeta- \varepsilon_{cr}|=\Delta_{so}/\tilde\Delta_{so}$ is calculated considering the spin-orbit interaction; $\Delta_{so}=13$ meV. The coordinate $\tilde p_{3,min}$ of the extremal orbit and $m^2(a')^2|B_2|/|\lambda|$ are calculated with Eqs.~(\ref{21}) and (\ref{38}), respectively. The values of the parameters $\beta^{\Sigma}$, $\lambda^{\Sigma}$ and $\beta^{\Delta}$, $\lambda^{\Delta}$ are found, taking into account Eqs.~(\ref{34}).}
\begin{tabular}{l|cccccc}
$|\zeta-\varepsilon_{cr}|_{i,0}$\ (meV)&19\ \ &21\ \ &24\ \ &27\ \ &36\ \ &58\ \ \\
\colrule
$\kappa$&1.46&1.62&1.85&2.08&2.77&4.5 \\
$\tilde\Delta_{so}$&0.53&0.50&0.45&0.42&0.33&0.21 \\
$|\zeta-\varepsilon_{cr}|_{i}$\ \ (meV)&24.5&26&28.9&31&39.4&62 \\
$\tilde p_{3,min}$&0.51&0.52&0.53&0.54&0.55&0.57\\
frequency $F$\ (T)&15.3&8.5&14&18.9&23&8.4\\
$\frac{m^2(a')^2|B_2|}{|\lambda|}$\ (eV)&1.9&7.8&4.2&3&4.8&80\\
\colrule
crossing point&$\Delta$&$\Sigma$&$\Delta$&$\Delta$&$\Delta$&$\Sigma$ \\
$m\beta$&2.3&1.5&3&2.9&3.7&4.9\\
$-\lambda\cdot10^3$&21&16&12&13&8&1.5\\
\end{tabular}
\end{table}

We now discuss the phases $\phi_i$ of the oscillations. Formula (\ref{33}) enables one to estimate the expected $\delta$ for the extremal orbits on the necks of the Fermi surface. With the values of $|\lambda|$, $\tilde p_{3,min}$, and $\tilde\Delta_{so}$ from Table II,  we find that $|\delta|$ is very close to $1/2$, and the difference $1/2-|\delta|$ does not exceed $0.005$. In other words, the effect of the spin-orbit interaction on the phase of the oscillations in ZrSiS is expected to be small, $\phi \approx 0$ for the extremal orbits near the points $\Delta$ and $\Sigma$, and the splitting of the oscillations should not be detected in this semimetal. The measured phases $\phi_2$  do lie in the vicinity of zero, Table I. However,  $\phi_1$ essentially differs from this zero value. Moreover, the value $\phi_1 \approx 0.53$ in Table I can hardly be attributed to the discussed change in the shape of the oscillations since this value is obtained with the low-field data (these data \cite{Hu1} give $n_0\approx 0.34$, and so  $\phi_1 \approx -n_0-1/8= -0.47$ or  $\phi_1 \approx 0.53$). Besides, a tendency to the splitting of the oscillations was clearly observed in Refs.~\cite{Hu1} and \cite{matus}, and the data of these papers lead to $\delta\approx 0.24$ (or $0.26$) and  $\delta\approx 0.1$ (or $0.4$), respectively. However, if $0 < \delta < 1/2$, the sharp minima or maxima of the quantum oscillations have to split, whereas the magnetization measurements \cite{Hu1} reveal the splitting of its {\it flat} maxima rather than its {\it sharp} minima. This fact and hence the obtained values of $\delta$ remain puzzling and require an additional experimental investigation.

\subsection{Extremal cross section passes through the point $\Sigma$}

The value $\phi_1\approx 0.53$ and a deviation of $\delta$ from zero or $1/2$ can be explained if we assume that $\tilde\Delta_{so}$ is close to its critical value $\tilde\Delta_c$ determined by Eq.~(\ref{24}), see Fig.~\ref{fig11}. In this case, the Fermi-surface neck is located either at the point $\Sigma$ (if $1> \tilde\Delta_{so}\ge \tilde\Delta_c$) or very close to it (if $\tilde\Delta_{so}\lesssim \tilde\Delta_c$). To evaluate possibility  of this situation, let us assume that  $\tilde\Delta_{so} =\tilde\Delta_c$, and hence the minimal cross section passes through the point $\Sigma$.

At $\tilde\Delta_{so}=\tilde\Delta_c$, the function $m_{*,min}(\tilde\Delta_{so})$ in the right hand side of Eq.~(37) is replaced by the function (\ref{26}) with $\tilde\Delta_{so}= \tilde\Delta_c$, whereas the left hand side of this equation can be rewritten as $|\zeta-\varepsilon_{cr}|_0/ |\zeta-\varepsilon_{cr}|$ by definition of $\kappa$ and $\tilde\Delta_{so}$. Hence, with this equation, $|\zeta-\varepsilon_{cr}|$ is expressible in terms of $|\zeta-\varepsilon_{cr}|_0$ and $\tilde\Delta_c$. In Eq.~(\ref{38}), the factor $F_3(\tilde p_{3,min},\tilde\Delta_{so})$ is now replaced by the function $F_3(0,\tilde\Delta_c)$ defined by formula (\ref{22}), and the estimate  $ma_{\Sigma}'|B_2^{\Sigma}|^{1/2} \approx 350$ (meV)$^{1/2}$ follows from Eqs.~(\ref{34}). Thus, at given value of $|\zeta-\varepsilon_{cr}|_0$, formula (\ref{38}) becomes the equation in $\tilde\Delta_c$,
\begin{eqnarray}\label{39}
 \frac{3\pi ma'|B_2|^{1/2}\mu_BF}{2|\zeta -\varepsilon_{cr}|_0^{3/2}}=
 \frac{[(1+\tilde\Delta_c)E(t) -\tilde\Delta_cK(t)]^{3/2}}{(1+ \tilde\Delta_c)[E(t)-\tilde\Delta_c K(t)]^{1/2}},
\end{eqnarray}
where $t=(1-\tilde\Delta_c)^{1/2}/(1+\tilde\Delta_c)^{1/2}$.
If $|\zeta-\varepsilon_{cr}|_0=58$ meV, see Table I, we eventually obtain the following values of the parameters:
\begin{eqnarray*}
 \tilde\Delta_c^{\Sigma}\!\!&\approx&\!\!0.75,\ \ \ \ \  |\zeta-\varepsilon_{cr}^{\Sigma}|\approx 165\, {\rm meV}, \\
 \lambda^{\Sigma}\!\! &\sim&\!\! -0.29,\ \ \  m\beta_{\Sigma}\sim 0.36,
   \end{eqnarray*}
where $\lambda^{\Sigma}$ and $m\beta_{\Sigma}$ have been calculated with formulas (\ref{24}) and (\ref{34}), respectively.
Note that for these values of the parameters to occur, the spin-orbit gap $2\Delta_{so}^{\Sigma}$ at the point $\Sigma$ has to be sufficiently large:
 \[
 2\Delta_{so}^{\Sigma}=2\Delta_c^{\Sigma}=2\tilde\Delta_c^{\Sigma} |\zeta-\varepsilon_{cr}^{\Sigma}|\approx 248\, {\rm meV}.
 \]
This gap is much larger than the measured one \cite{uykur} and than  the gap $\sim 20$ meV obtained in the band-structure calculations \cite{schoop}. As to the chemical potential measured from the edge of the electron energy band, it retains the modest value,
 \[
 |\zeta-\varepsilon_{cr}^{\Sigma}|-\!\Delta_{so}^{\Sigma} \approx 41\, {\rm meV}.
 \]
With decreasing $|\zeta-\varepsilon_{cr}|_0$, the required spin-orbit gap $2\Delta_{so}^{\Sigma}$ increases, and at $|\zeta-\varepsilon_{cr}|_0=21$ meV, we find the unrealistic value $2\Delta_{so}^{\Sigma}\sim 2.25$ eV although $\lambda^{\Sigma}\approx 0.12$ and $|\zeta-\varepsilon_{cr}^{\Sigma}|-\!\Delta_{so}^{\Sigma} \approx 14$ meV. Thus, the obtained estimates of $2\Delta_{so}^{\Sigma}$ argue against the assumption that $\tilde\Delta_{so}\approx \tilde\Delta_c$. However, it is worth noting that the exact values of the cyclotron mass $m_{*,1}$  determining $|\zeta-\varepsilon_{cr}|_{1,0}$ and of $ma_{\Sigma}' |B_2^{\Sigma}|^{1/2}$ are crucial in verifying this assumption.

\subsection{Both extremal cross sections are near the point $\Sigma$}

The data of Ref.~\cite{matus} lead to the close values of $|\zeta-\varepsilon_{cr}|_{i,0}$ for the frequencies $F_1$ and $F_2$, see Table I. This might indicate that the appropriate cross sections correspond to one and the same crossing point, i.e., the frequency $F_1$ refers to  the extremal orbit on the neck of the Fermi surface near the point $\Sigma$ whereas $F_2$ is produced by the central cross section containing this point. Although this interpretation, as was mentioned above, contradicts the results of Refs.~\cite{wang1,singha}, we consider it here to demonstrate how formulas (\ref{10})  and (\ref{23}) can work. Under this interpretation, the ratio $F_1/F_2\approx 0.56$ enables one to find the parameter $\lambda$. Neglecting the spin-orbit interaction, we find from Eq.~(\ref{10}):
 \begin{equation*}
|\lambda^{\Sigma}|\approx \frac{3}{\pi^2}\frac{F_1^2}{F_2^2}\approx 0.095.
 \end{equation*}
If one takes into account the spin-orbit interaction, the right hand side of this formula, according to Eq.~(\ref{23}), has to be divided by $[f(\tilde \Delta_{so})]^2$ where $\tilde \Delta_{so}\approx 0.5\div 0.53$ and $f(\tilde \Delta_{so})\approx 1.25$, see Fig.~\ref{fig8}. In other words, with the spin-orbit coupling, we find $\lambda^{\Sigma}\approx -0.061$. However, this value of $\lambda$ essentially differs from the estimate $\lambda^{\Sigma} \approx -0.016$ obtained above for the same crossing point. This discrepancy is an added reason for ascribing the frequencies $F_1$ and $F_2$ to the different crossing points.

Recently a fresh interpretation of the oscillation frequencies was presented by M\"uller et al.\ \cite{muller}. They measured the de Haas - van Alphen oscillation in ZrSiS and, apart from the frequency $F_1\approx 8$ T, detected the {\it two} frequencies $F_{2a}\approx 16$ T and $F_{2b}\approx 22$ T instead of the single frequency $F_2$. M\"uller et al.\ ascribed the frequency $F_1$ to the extremal orbit on the neck near the point $\Sigma$, $F_{2a}$ to the central cross section containing this point, and $F_{2b}$ to the extremal orbit on the neck near the point $\Delta$. Note that the existence of the two frequencies $F_{2a}$ and $F_{2b}$ permits one to avoid the above-mentioned contradiction with the results of the papers \cite{wang1,singha}, assuming that the phase of the oscillations in these papers was measured for the frequency corresponding to the orbit on the neck. Interestingly, with Eq.~(\ref{11}) and the values of the cyclotron masses measured in Ref.~\cite{muller}, we find that  $|\zeta-\varepsilon_{cr}|_0 \approx 19.9$, $23.2$, $20.1$ meV for the cases of the frequencies $F_1$, $F_{2a}$, $F_{2b}$, respectively. Since $|\zeta-\varepsilon_{cr}|_0$ for $F_1$ and $F_{2b}$ are very close to each other, we assume here that just the frequency $F_{2b}$ (rather than $F_{2a}$) corresponds to the central cross section. Then, using Eq.~(\ref{10}) with $F_1/F_{2b}\approx 0.36$ and taking into account the spin-orbit interaction, we arrive at the estimate $\lambda^{\Sigma}\approx -0.26$ which is in reasonable agreement with $\lambda^{\Sigma}\approx -0.16$, considering the approximate character of the values in formulas (\ref{34}).

\section{Conclusions}

We analyze a crossing of two band-contact lines in the Brillouin zones of crystals. In the vicinity of the crossing point of such lines, the electron spectrum essentially differs from the Dirac spectrum occurring in the planes perpendicular to an isolated band-contact line. Taking into account this difference, we theoretically investigate the possible types of the Fermi surface and its characteristics near the crossing point. We calculate the quantities commonly measured in the quantum-oscillation experiments, viz., the extremal cross-sectional areas $S_{ex}$, the cyclotron masses $m_*$, and the phase of the oscillations $\phi$. The especial emphasis in our analysis is given to the case of the nodal-line semimetals for which the dispersion of the contacting bands along the nodal lines is small as compared to the ordinary scale of the electron band structure. In this situation  the appropriate formulas are essentially simplified. We calculate the Fermi-surface characteristics both without and with considering the weak spin-orbit interaction. This interaction introduces only quantitative corrections to the cross-sectional areas and cyclotron masses found in neglect of the interaction. For the orbits the planes of which are sufficiently far from the crossing point, the spin-orbit interaction has no effect on the phase of the oscillations $\phi$ that is still equal to zero and is specified by the Berry phase ($\Phi_B=\pi$) in absence of this interaction. However, for the extremal  orbits near the crossing point, the spin-orbit interaction can noticeably change the phase $\phi$.

To illustrate the obtained results, we apply them to ZrSiS in which the crossing of the nodal lines occurs. We analyze several possible interpretations of the experimental data obtained in  Refs.~\cite{Hu1,Ali1,wang1,singha,matus} and within these interpretations, estimate the parameters of the spectrum and specifically the position of the chemical potential relative to the crossing-point energies. Our analysis shows that the quantum oscillations with the low frequencies $F_1$ and $F_2$, Table I, seem to be produced by the extremal orbits lying on the necks of the Fermi surface near the points $\Sigma$ and $\Delta$, see Fig.~\ref{fig1}. However, this analysis does not permit us to interpret all the experimental data unambiguously since the certain data concerning the quantum oscillations of the frequency $F_1$ remain puzzling. These oscillations require further experimental investigations.


\begin{thebibliography}{}
\bibitem{com1} For the fourfold symmetry axis when $B_1=B_2$ and $B_1'=B_2'$, the $E_{c,v}({\bf p})$ can contain the term $\beta^2(p_1^2-p_2^2)^2$ instead of $\beta^2p_1^2p_2^2$. However, this case reduces to Eq.~(\ref{2}) by the rotation of the cooordinate axes $p_1$, $p_2$ by the angle $\pi/4$.
\bibitem{com2} The self-intersection of the Fermi surface is absent in the line S -- $\Sigma$ since $\zeta-\varepsilon_{cr}^S> \varepsilon_{cr}^{\Sigma}-\varepsilon_{cr}^S$.
\bibitem{com} Note that the value of $B_2'$ is sufficiently large for the point  $\Delta$ since the band-contact line is essentially curved near this point in the $p_1$-$p_2$ plane \cite{pez,fu19}.
\bibitem{com3} If we assume that the dispersion law along the nodal line between the crossing points $i$ and $j$ has the form: $\varepsilon(p)= 0.5(\varepsilon_{cr}^j+ \varepsilon_{cr}^i)- 0.5(\varepsilon_{cr}^j- \varepsilon_{cr}^i)\cos(\pi p/p_{ij})$, the appropriate $B_n$ at the point $i$ will be $(\pi^2/4)(\varepsilon_{cr}^j- \varepsilon_{cr}^i)/p_{ij}^2$, i.e., in $\pi^2/4$ larger than that given in the text. Then, $|\lambda|$ increases by $\pi^4/16$ times.
\end{thebibliography}
\end{document}